\documentstyle[11pt,amscd]{article}                                    
\newcommand{\frak}[1]{{\mathbf #1}}                                  

\def\AFOUR{%
\setlength{\textheight}{9.0in}%
\setlength{\textwidth}{5.75in}%
\setlength{\topmargin}{-0.375in}%
\hoffset -.5in%
\renewcommand{\baselinestretch}{1.17}%
\setlength{\parskip}{6pt plus 2pt}}%
\AFOUR                                          
\def\car{\mathop{\square}}
\def\carre#1#2{\raise 2pt\hbox{$\scriptstyle #1$}\car_{#2}}
\makeatletter
\def\section{\@startsection {section}{1}{\z@}{-3.5ex plus -1ex minus
 -.2ex}{2.3ex plus .2ex}{\large\bf}}
\def\subsection{\@startsection{subsection}{2}{\z@}{-3.25ex plus -1ex 
minus
 -.2ex}{1.5ex plus .2ex}{\normalsize\bf}}
\makeatother
\newcommand{\nc}{\newcommand}
\newcommand{\rnc}{\renewcommand}
\nc{\be}{\begin{equation}}
\nc{\ee}{\end{equation}}
\nc{\bea}{\begin{eqnarray}}
\nc{\eea}{\end{eqnarray}}

\def\href#1#2{{#2}}

\rnc{\a}{\alpha}
\nc{\ab}{\bar{\a}}
\nc{\ap}{\a^{+}}
\nc{\abm}{\ab^{-}}
\rnc{\b}{\beta}
\nc{\bb}{\bar{\b}}
\nc{\bbp}{\bb_{\zb}^{+}}
\nc{\bm}{\b_{z}^{-}}
\nc{\oa}{\overline{\a}}
\nc{\ob}{\overline{\b}}
\rnc{\gg}{\gamma}
\rnc{\d}{\delta}
\nc{\f}{\phi}
\nc{\fb}{\bar{\phi}}
\nc{\vf}{\varphi}
\nc{\p}{\psi}

\rnc{\c}{\chi}
\nc{\la}{\lambda}
\nc{\m}{{\mathrm m}}
\nc{\n}{\nu}
\rnc{\o}{\omega}
\nc{\Om}{\Omega}
\rnc{\t}{\theta}
\nc{\eps}{\epsilon}
\rnc{\S}{\Sigma}
\nc{\F}{\Phi}
\nc{\trac}[2]{{\textstyle\frac{#1}{#2}}}
\nc{\ex}[1]{\mbox{e}^{\,\textstyle#1}}
\nc{\mat}[4]{\left(\begin{array}{cc}#1&#2\\#3&#4\end{array}\right)}
\nc{\som}[9]{\left(\begin{array}{ccc}#1&#2&#3\\#4&#5&#6\\#7&#8&#9%
\end{array}\right)}
\nc{\tr}{\mathop{\mbox{tr}}\nolimits}
\nc{\ad}{\mathop{\mbox{ad}}\nolimits}
\nc{\Tr}{\mathop{\mbox{Tr}}\nolimits}
\nc{\Det}{\mathop{\mbox{Det}}\nolimits}
\nc{\rk}{\mathop{\mbox{rk}}\nolimits}
\nc{\ra}{\rightarrow}
\nc{\Ra}{\Rightarrow}
\nc{\LRa}{\Leftrightarrow}
\nc{\ot}{\otimes}
\rnc{\ss}{\subset}
\nc{\nul}{\noindent\underline}
\nc{\non}{\nonumber\\}
\nc{\subs}[1]{{\vspace*{0.5cm}}%
{\noindent\underline{#1}}{\addcontentsline{toc}{subsection}{#1}}%
{\vspace*{0.3cm}}}
\nc{\zb}{\bar{z}}
\rnc{\lg}{\frak{g}}
\nc{\lt}{\frak{t}}
\nc{\lk}{\frak{k}}
\nc{\lh}{\frak{h}}
\nc{\pik}{\Pi_{\lk}}
\nc{\pip}{\Pi_{+}}
\nc{\pim}{\Pi_{-}}
\nc{\pih}{\Pi_{\lh}}
\nc{\jz}{J_{z}}
\nc{\jzh}{\jz^{\lh}}
\nc{\jzp}{\jz^{+}}
\nc{\jzm}{\jz^{-}}
\nc{\del}{\partial}
\nc{\dz}{\del_{z}}
\nc{\dzb}{\del_{\bar{z}}}
\nc{\az}{A_{z}}
\nc{\azb}{A_{\bar{z}}}
\nc{\g}{g^{-1}}
\nc{\dw}{\Delta_{W}}
\nc{\Ad}{{\mbox{Ad}}}
\nc{\ks}{Ka\-za\-ma-\-Su\-zu\-ki}
\nc{\KS}{\ks}
\nc{\ksm}{\ks\ model}
\rnc{\AA}{{\Bbb A}}
\nc{\BB}{{\Bbb B}}
\nc{\CC}{{\Bbb C}}
\nc{\PP}{{\Bbb P}}
\nc{\cpm}{\CC\PP(m)}
\nc{\cpn}{\CC\PP(n)}
\nc{\cp}[1]{\CC\PP(#1)}
\nc{\gmn}{G(m,m+n)}
\nc{\gmnk}{\gmn_{k}}
\nc{\cO}{{\cal O}}
\nc{\bcO}{\bar{\cO}}
\nc{\bO}{\bar{O}}
\nc{\oQ}{\overline{Q}}
\nc{\ie}{{\it i.e.~}}
\nc{\eg}{{\it e.g.~}}
\begin{document}
\global\parskip 4pt
\makeatother\begin{titlepage}
\begin{flushright}
{ROM2F/99-42}
\end{flushright}
\vspace*{0.5in}
\begin{center}
{\LARGE\sc  Discrete torsion in non-geometric orbifolds}
\vskip 0.5cm
{\LARGE\sc and their open-string descendants}\\
\vskip 0.8cm
\makeatletter

\begin{tabular}{c}
{\bf Massimo Bianchi\footnotemark, Jos\`e F. Morales\footnotemark,
Gianfranco Pradisi\footnotemark }
\\
Dipartimento di Fisica \\
Universit\`a degli studi di Roma ``Tor Vergata''\\
and
INFN, sezione di Roma ``Tor Vergata'' \\
Via della Ricerca Scientifica, 1 - 00173 Rome, ITALY\\
\end{tabular}

\end{center}
\addtocounter{footnote}{-2}%
\footnotetext{e-mail: massimo.bianchi@roma2.infn.it}
\addtocounter{footnote}{1}%
\footnotetext{e-mail: morales@roma2.infn.it}
\addtocounter{footnote}{1}%
\footnotetext{e-mail: pradisi@roma2.infn.it}
\vskip .50in
\begin{abstract}
\noindent
We discuss some $Z_N^L\times Z_N^R$ orbifold compactifications of 
the type IIB superstring to $D= 4,6$ dimensions and their type I 
descendants.
Although the $Z_N^L\times Z_N^R$ generators act asymmetrically on the
chiral string modes, they result into left-right symmetric 
models that admit sensible unorientable reductions.   
We carefully work out the phases that appear in the modular
transformations of the chiral amplitudes and identify the possibility 
of introducing discrete torsion.
We propose a simplifying ansatz for the construction of the 
open-string
descendants in which the transverse-channel Klein-bottle, annulus and
M\"obius-strip amplitudes are numerically 
identical in the proper parametrization
of the world-sheet. A simple variant of the ansatz for the
$Z_2^L\times Z_2^R$ orbifold gives rise to models
with supersymmetry breaking in the open-string sector.

\end{abstract}
\makeatother
\end{titlepage}
\begin{small}
\end{small}

\setcounter{footnote}{0}

\section{Introduction}

The microscopic description of Bogomolny-Prasad-Sommerfield (BPS) 
solitons
carrying Ramond--Ramond (R-R) charge in terms of
Dirichlet-branes (D-branes) and Orientifold-planes (O-planes) 
\cite{cjp}
has played a crucial r\^ole
in the emerging non-perturbative picture of string theory.
Although many interesting  vacuum configurations of the type II
superstring and their type I descendants can be easily accounted 
for in terms of D-branes and O-planes,
these concepts tend to loose their clear meaning in non-trivial
compactifications, such as asymmetric orbifolds
\cite{nsv}, free fermionic models \cite{fermi, mbas} and Gepner 
models
\cite{gep, gepner}. Insisting on a `geometric' target-space approach, that is 
expected to be valid in the large volume limit,
is by far less useful than pursuing an `algebraic' worldsheet approach
based on the conformal field
theory (CFT) description \cite{pss, fucsch, recsch}.
The construction of type I descendants of  
non-geometric type II vacuum configurations 
may shed some light on the necessary generalization of the 
above concepts.

It is the purpose of this paper to discuss some
$Z_N^L\times Z_N^R$ models that, though left-right symmetric, do not 
admit a clear
geometric description because of the chiral nature of the projections.
Still, we are able to analyze their open-string descendants in
terms of two-dimensional CFT's where the strict concepts of D-branes 
and O-planes are abandoned.
Differently from what has been found for geometric orbifolds, we will
find the presence of open strings belonging to twisted sectors
of the orbifold group.
Similar kinds of open strings have been recently found
in a discussion of type I vacuum configurations in $D=6$ that 
involve
D-branes at angles~\cite{blgo}.
We cannot exclude the possibility that some
of the models discussed in the present paper may admit a geometric 
interpretation in terms of
M-theory along the lines of~\cite{drs}.

We will identify the possibility of introducing
additional phases (known as discrete torsion \cite{distor}) in the 
modular invariant
combinations of characters that appear in the one-loop torus 
amplitude. Some brane configurations in the presence of discrete torsion 
have been considered in \cite{koushik}.
The introduction of discrete torsion allows one to relate models with
different amount of supersymmetry.
The resulting theories are symmetric under the
interchange of the left and right movers for any choice of the
discrete torsion and are thus good candidate
parents of type I descendants. The very possibility of working in a 
rational context (only a highly
restricted class of lattices admit chiral automorphisms! \cite{lsw}) 
makes
the construction of the type I
descendants almost straightforward following the approach pioneered by
Sagnotti \cite{ascar} and then systematized in \cite{mbas,
pss, gpas, bpstor}.

In this paper we are able to go a step further and simplify the
construction of at least one of the possible descendants by assuming 
that
the transverse-channel Klein-bottle ($\widetilde{\cal K}$), Annulus 
($\widetilde{\cal A}$) and M\"obius-strip ($\widetilde{\cal M}$) 
amplitudes exactly coincide in the proper parametrization of the 
worldsheet.
The numerical relation $\widetilde{\cal K}=
\widetilde{\cal A}=\widetilde{\cal M}$
automatically enforces the tadpole conditions.
With this simplifying ansatz, the only non-trivial issue consists in
reconstructing the open-string
spectrum in terms of Chan-Paton (CP) group assignments
from the overall transverse-channel multiplicities.

In the transverse channel, the Klein-bottle amplitude $\widetilde{\cal K}$ 
represents
a closed-string exchange between ``crosscap-states'' $|C\rangle$ whose
target-space counterparts are the loci left invariant under the
combined action of orientation reversal ($\Omega$) and some 
target-space symmetry (${\cal I}$).   
Although these objects are well defined in a CFT context, they
do not necessarily admit a sensible large volume limit.
They share with the standard O-planes the property of
being charged with respect to the R-R fields. The cancellation of the
R-R charge flowing through the compact space requires as usual the 
introduction of ``boundary-states'' $|B\rangle$ carrying opposite R-R charges.
A ``minimal choice'' is provided by boundaries $|B\rangle$ with the
same closed-string content as $|C\rangle$, where the difference
in the R-R charges between the two objects is compensated by a correct
assignment of CP multiplicities.
We will consider this minimal ansatz, according to which,
once the CP multiplicities have been plugged into the annulus amplitude 
$\widetilde{\cal A}$,
the $\langle B|B\rangle$ exchange numerically coincides
with the $\langle C|C\rangle$ exchange 
even for the whole tower of massive states.
The M\"obius-strip amplitude $\langle B|C\rangle+\langle C|B \rangle$,
when expressed in terms of
``hatted'' quantities \cite{mbas} as required by the reality of 
$\widetilde{\cal M}$,
will again coincide with the transverse annulus and Klein-bottle
amplitudes, up to alternating signs at the massive levels.
The construction is not as restrictive as one could imagine. One may
easily check that in the case of toroidal orbifolds it encompasses most of
the $Z_2,Z_3$ orientifold models  
considered in the recent literature \cite{gpdpgjbz}.
A general list of solutions, and a rather more precise conformal
description of the models we consider, can be systematically found
following the open-descendant techniques developed in \cite{gpas, 
mbas, bpstor, pss}.
Indeed, it is explicitly shown in Appendix B that the 
$Z_2^L\times Z_2^R$ model
provides an orientifold description of the open-string 
descendant associated
to the $A_{16}$ permutation invariant considered in \cite{gepner}.   
The same is true for the $Z^L_3\times Z^R_3$ model in 
$D=6$ indicated as $A_{81}$ in \cite{gepner}, although we omit
here the details of a similar correspondence. 

Some circumstantial evidence for the validity of our simplifying 
ansatz can be illustrated in the
simplest context of the so-called toroidal orientifold models 
\cite{pw}.
These models are obtained by quotienting the type II superstrings 
by $\Omega {\cal I}$, where ${\cal I}$ is the inversion
of the (internal) coordinates of a $d$-dimensional torus.
They are T-dual to standard toroidal compactifications of the type
I superstring in the presence of Wilson lines breaking $SO(32)$
to $SO(2^{5-d})^{2^{d}}$ \cite{bpstor}.
In the absence of Wilson lines the relation
 $\widetilde{\cal K}=
\widetilde{\cal A}=\widetilde{\cal M}$ does not hold
because crosscap states only couple to even windings while boundary
states couple to all windings.
Precisely after introducing the proper Wilson lines \cite{bpstor, 
cjp, pw} the
relation  $\widetilde{\cal K}=
\widetilde{\cal A}=\widetilde{\cal M}$
is enforced. All the open-string KK momenta are shifted by one-half unit
and the transverse-channel annulus only allows even windings to flow.

Slightly at variant with the minimal ansatz, in the last part of the 
paper we construct a non-supersymmetric R-R tadpole-free model. 
The prize to pay
is an uncancelled Neveu-Schwarz--Neveu-Schwarz (NS-NS) tadpole.

It is amusing to observe that in the absence of discrete torsion
untwisted and twisted open-string states combine with one another  
to reconstruct multiplets of the enhanced bulk supersymmetry.  This 
could sound surprising since twisted open-strings are naturally 
interpreted as strings connecting ``branes at 
angles'' \cite{blgo}, each of them breaking one-half of the original 
$32$ type IIB supercharges. However in the present context the two
objects are properly identified by elements of the orbifold group and 
the distinction between them tends to loose meaning.
In the $Z_2^L\times Z_2^R$ case this is in line with the 
identification of one of the two chiral projections as the 
T-duality group element that inverts all the radii.

The plan of the paper is as follows.
In Section 2 we briefly review the construction of $Z_N^L\times 
Z_N^R$ 
asymmetric orbifolds of the type IIB superstring.
Section 3 is devoted to the explicit construction of the 
$Z_2^L\times Z_2^R$
orbifold in $D= 6$ in the presence of discrete torsion
and to the analysis of its type I descendants.
We partly associate the rank reduction of the CP
group to the presence of a quantized NS-NS antisymmetric tensor 
background
\cite{bpstor, mbtor} and partly to the identification of 
would-be D5- and D9-branes imposed after quotienting by T-duality.
In Sections 4 and 5 we perform similar analyses for
$Z_3^L\times Z_3^R$ orbifolds in $D= 4,6$ and discuss their
type I descendants. In Section 6
we describe a variant of the $Z_2^L\times Z_2^R$ model in $D= 6$
that leads to brane supersymmetry breaking \cite{bach, penta, bsb, iban}.
Finally, Section 7 contains our conclusions and comments for future 
developments of the present approach 
to other non-geometric vacuum configurations that admit simple and 
handy
algebraic descriptions \cite{gepner, recsch, fucsch}.
In order to make the paper as self-contained as possible,
we have added two appendices. In Appendix A we set up our conventions 
and define the conformal blocks that appear in the 
asymmetric orbifolds under consideration. In Appendix B we have included an
expansion of the $Z_2^L\times Z_2^R$ 
models in terms of generalized characters. 

\section{Asymmetric orbifolds: a quick review}

Thanks to the large degree of independence between left and right 
movers
on the string worldsheet, one can conceive vacuum configurations
in which one set of modes propagates in a target space and the other
in a completely different one. For closed strings, modular invariance
puts very tight constraints and the largest class of models of this
kind that have been constructed are the asymmetric orbifolds of 
Narain,
Sarmadi and Vafa \cite{nsv}. Free fermionic \cite{fermi}
and covariant lattice \cite{lsw} constructions have some overlap 
with asymmetric orbifolds.

An asymmetric orbifold is obtained by quotienting a string
compactification, typically on a $d$-dimensional torus, by a discrete
group, typically a cyclic group, that acts asymmetrically on the left
and right movers. For the heterotic string, 
that is left-right asymmetric from the very beginning, 
this is rather natural. For the type II
superstrings, in particular for the type IIB superstring 
that is left-right
symmetric, this may sound slightly artificial, but undoubtedly
represents an improvement and an implementation of the string
potentialities beyond their field-theory limit \cite{dkv}.
Most of these asymmetric constructions are still waiting for
some geometric interpretation in terms of M-theory or F-theory, if any. 

We will concentrate most of our analysis on
$Z_N^{L} \times Z_{N}^{R}$ asymmetric orbifolds of toroidal
compactifications of the type IIB superstring\footnote{The
$Z_3^{L} \times Z_{3}^{R}$ case in $D= 4$ has
been previously considered in \cite{dabhar}.
We will find results somewhat in disagreement with \cite{dabhar}
for the oriented closed-string spectra.} and their
open-string descendants.

The basic building blocks in the construction of $Z_N$ asymmetric
orbifolds, are the chiral supertraces
\be
\rho_{g,h}\equiv {\rm Tr}^\prime_{\rm NS,g} \frac{1}{2}(1-(-)^{F})h
q^{L_0-\frac{c}{24}}- {\rm Tr}^\prime_{\rm R,g}\frac{1}{2}(1+(-)^{F})h
q^{L_0-\frac{c}{24}} \quad ,
\label{rhogh}
\ee
where $g,h\in Z_N$ are elements of the chiral orbifold
group and the trace runs over the 
$g$-twisted sector with a plus for NS states and minus for R states. 
We denote by a prime the omission in the trace
(\ref{rhogh}) of the free bosonic contributions
\be
X_D =  (\sqrt{\tau_2}\eta\bar{\eta})^{2-D}
\ee
for non-compact bosons, and
\be
\Lambda_\Gamma = \frac{1}{(\eta\bar{\eta})^{d}}
\sum_{{\bf p}\in \Gamma_{d,d}} q^{\frac{1}{2}{\bf p}_L^2}
\bar{q}^{\frac{1}{2}{\bf p}_R^2}
\ee
for compact bosons, with $D+d= 10$.
In addition, string partition functions will be weighted by the volume
factors
\bea
{\cal V}_D&\equiv& \tau_2^{\frac{D}{2}}\int \frac {d{\bf p}\,d{\bf x}}
{(2\pi)^D}
e^{-\pi\tau_2 \alpha^\prime{\bf p}^2}= \frac{V_D}
{(4\pi^2\alpha^\prime)^{\frac{D}{2}}}
\eea
and by an integer $C_{g,h}$ counting the number of ``fixed points''
under a given $g$-action left invariant by the $h$-projection.
It can be computed applying the formula \cite{nsv}
\be
C_{g,h}= \left|\frac{N_g}{(1-g)N_g^*+(1-h)N_g}\right|
\label{fixedpoints} \quad ,
\ee
where $N_g$ represents the lattice orthogonal to the lattice
left invariant by $g$ and $N_g^*$ its dual.
Notice that a lattice admitting a chiral discrete
automorphism must be very special.
The choice is typically restricted to weight lattices of
compact Lie algebras \cite{lsw}.

\section{$T^4/Z_2^L\times Z_2^R$ orbifold and
its open-string descendants}

Let us start with the $T^4/Z_2^L\times Z_2^R$ orbifold of the type
IIB superstring. As a choice for a $T^{4}$ admitting a chiral $Z_2$
isometry we take the torus associated to the weight lattice of 
$SO(8)$.
For future reference, notice that this requires to turn on a quantized
NS-NS antisymmetric tensor background of rank two. The construction
is equivalent to a T-duality orbifold of the standard geometric
$Z_{2}$ orbifold. Orbifolding by T-duality has been considered in
\cite{dabhar} and \cite{dinsil} as a way to freeze out some of the
moduli of the theory. From the open-string perspective it has been
considered in \cite{carlo}.

The torus partition function can be written as
\be
{\cal T}= {\cal V}_6\int_{\cal F}\frac{d^2 \tau}{\tau_2^2}\,X_6
\sum_{g_L,g_R}
{\cal T}_{g_L,g_R} \quad ,
\label{torus6}
\ee
where ${\cal F}$ is the fundamental region of the one-loop moduli
space and 
\bea
{\cal T}_{00}&= &\frac{1}{4}\left[ \rho_{00}\bar\rho_{00}
\Lambda_{SO(8)}+\rho_{00}\bar\rho_{01}\Lambda_{R}
+ \rho_{01}\bar{\rho}_{00}\bar{\Lambda}_{R}
+\rho_{01}\bar{\rho}_{01}\right]\label{t00} \nonumber \\
{\cal T}_{01}&= &\frac{2}{4}\left[ \rho_{00}\bar{\rho}_{10}
\Lambda^+_{W}+\rho_{00}\bar{\rho}_{11}\Lambda^-_{W}
+ \epsilon\rho_{01}\bar{\rho}_{10}
+\epsilon\rho_{01}\bar{\rho}_{11}\right]\label{t01}\nonumber \\
{\cal T}_{10}&= &\frac{2}{4}\left[ \rho_{10}\bar{\rho}_{00}
\Lambda^+_{W}+\rho_{11}\bar{\rho}_{00}\Lambda^-_{W}
+ \epsilon\rho_{10}\bar{\rho}_{01}
+\epsilon\rho_{11}\bar{\rho}_{01}\right]\label{t10}\nonumber \\
{\cal T}_{11}&= &\frac{16}{4}\left[ \rho_{10}\bar{\rho}_{10}
+\rho_{11}\bar{\rho}_{11}
-\frac{\epsilon}{2}\rho_{10}\bar{\rho}_{11}
-\frac{\epsilon}{2}\rho_{11}\bar{\rho}_{10}\right] \quad .
\label{t11}
\eea
The explicit form of the chiral supertraces $\rho_{gh}$ and 
lattice sums is given in Appendix A.
The relative powers of two represent the number of ``fixed
points'' under the asymmetric orbifold group actions.
Most of the amplitudes in
(\ref{t11}) are in the same modular orbit as
the amplitudes in the untwisted sector and therefore the relevant number
of fixed points (associated to $C_{g,1}$) is easily determined
from modular transformations. This is not the case for
the modular orbit ${(g_L,h_R)}$ which is clearly disconnected from the
untwisted sector. By inspection of (\ref{fixedpoints}) one can easily
see that $C_{g_L,h_R}= C_{g_L,1}$ and therefore the multiplicity of this
orbit is again determined through modular transformations
up to a $Z_2$-phase, \ie a sign. This phase, that we have 
denoted by $\epsilon$ in (\ref{t11}), 
represents the discrete torsion between the two $Z_2$ factors 
\cite{distor}.

Depending on the
choice of discrete torsion, $\epsilon= \pm 1$,
the spectrum of massless oriented closed-string states corresponds
to compactifications of the type IIB theory on spaces topologically 
equivalent
to $T^{4}$ and $K3$, respectively.
The resulting massless spectra of closed oriented strings are:\\  \\
\begin{tabular}{lll }
$\epsilon$ & Supersymmetry & Supermultiplets \\
  $ +$ & ${\cal N}= (2,2)$ & ${\bf G}_{(2,2)}$ \\
  $ -$ & ${\cal N}= (2,0)$ & ${\bf G}_{(2,0)}+21\, {\bf T}_{(2,0)}$\\
\end{tabular}\\ \\
where ${\bf G}$ and ${\bf T}$ stand for gravity and tensor 
super-multiplets
respectively.

The Klein-bottle amplitude is defined by the $\Omega$-projection of (\ref{t11}).
Clearly, left-right asymmetric sectors, such as ${\cal T}_{01}$ or 
${\cal
T}_{10}$, will not contribute to this trace since states in these 
sectors
come always in $\Omega$ even-odd pairs. The
result only involves the 
diagonal components $\delta_{g_1,g_2}\rho_{g_1,h_1-h_2}$ and reads
\be
{\cal K}= \frac{{\cal V}_6}{2}
\int_0^\infty \frac{d \tau_2}{\tau_2^4}\frac{1}{\eta^4}\left[
\frac{1}{2}\rho_{00}\Lambda^+_W+\frac{1}{2}\rho_{01}+ 
2^{-\frac{r}{2}}\, 8\, \rho_{10}
-\epsilon \, 2^{-{\frac{r}{2}}}\, 4 \, \rho_{11}\right](2i\tau_2) \quad ,
\label{klein}
\ee
where we have used the fact that the total number of $\pm$
$\Omega$-eigenvalues in the twisted-sector ground-states is given by
\cite{carlo,kaku}
\be
n_{\pm}= \frac{n_F}{2}(1\pm 2^{-\frac{r}{2}}) \quad ,
\ee
with $r= 2$ the rank of the antisymmetric tensor background $B_{ij}$ 
in the $SO(8)$ lattice and $n_F$ the number of fixed points
$C_{g,h}$ in the parent torus amplitude. Notice that this mechanism is
automatic in terms of characters, \ie $n_{+}-n_{-}$ characters
appear diagonally in the one-loop modular invariant (see Appendix B).

As usual NS-NS (R-R) states flowing along the Klein-bottle (\ref{klein})
are (anti)symmetrized, while
one half of the remaining ones survives the $\Omega$-projection.
The resulting spectra of massless unoriented closed-string states
are given by:\\ \\
\begin{tabular}{lll }
$\epsilon$ & Supersymmetry & Supermultiplets \\
  $ +$ & ${\cal N}= (1,1)$ & ${\bf G}_{(1,1)}+4\, {\bf 
V}^{c}_{(1,1)}$ \\
  $ -$ & ${\cal N}= (1,0)$ & ${\bf G}_{(1,0)}+14\,{\bf H}_{(1,0)}+7\,
  {\bf T}_{(1,0)}$\\
\end{tabular}\\ \\
In order to determine the unoriented open-string spectrum, 
that one has
to couple to the above unoriented closed-string spectrum,
we start by rewriting the Klein-bottle amplitude
(\ref{klein}) in the transverse channel\footnote{Although tilded and 
untilded
amplitudes coincide, we distinguish by a tilde the
rewriting of the one-loop amplitudes in terms of the closed-string
modular parameter of the common world-sheet double-cover.}
($\tau_2\rightarrow 1/\tau_2$) as
\be
{\widetilde{\cal K}}= 2^3\frac{{\cal V}_6}{2}
\int_0^1 \frac{d q}{2\pi q}\frac{1}{\eta^4}\left[
\rho_{00}\Lambda_R+ \rho_{01}
+ 2\rho_{10}+2\epsilon \rho_{11}\right](q) \quad .
\label{kleint}
\ee
According to our simplifying ansatz, the transverse-channel
annulus and M\"obius-strip amplitudes read
\bea
{\widetilde{\cal A}}&= &2^{-3}\frac{{\cal V}_6}{2}
\int_0^1 \frac{d q}{2\pi q}\frac{1}{4\eta^4}\left[
I_{O}^2(\rho_{00}O+\rho_{01}+2\rho_{10}+2\epsilon\rho_{11})
\right.\nonumber\\
&&\left.+\rho_{00}(I_V^2\, V+I_S^2\, S+
I_C^2\, C)+2(\rho_{10}-\epsilon\rho_{11})(I_V^2+I_S^2+I_C^2)\right](q)
\nonumber\\
{\widetilde{\cal M}}&= &-2\frac{{\cal V}_6}{2}
\int_0^1 \frac{d q}{2\pi q}\frac{I_O}{2\eta^4}\left[\rho_{00}O+\rho_{01}
+2\rho_{11}+2\epsilon\rho_{10}\right](-q) \quad ,
\eea 
with $I_{O}= 16$, $I_{V}= I_{S}= I_{C}= 0$. A simple CP group 
assignment
of the boundary traces is given by
\bea
I_{0}&= &n_1+n_2+n_3+n_4\nonumber\\
I_{V}&= &n_1+n_2-n_3-n_4\nonumber\\
I_{S}&= &n_1-n_2+n_3-n_4\nonumber\\
I_{C}&= &n_1-n_2-n_3+n_4 \quad ,
\eea
with $n_1= n_2= n_3= n_4= 4$.
Going to the direct-channel Annulus and M\"obius-strip 
amplitudes through $S$ and ${P}$ modular transformations respectively 
yields
\bea
{\cal A}&= &\frac{{\cal V}_6}{2}
\int_0^\infty \frac{d \tau_2}{\tau_2^4}\frac{1}{\eta^4}\left[
\sum_i n_i^2 [\frac{1}{2}\rho_{00} O+\frac{1}{2}\rho_{01}
+\rho_{10}+\epsilon\rho_{11}]+\sum_{i<j} 2 n_i n_j(\rho_{10}-\epsilon
\rho_{11})\nonumber
\right.\\
&&\left.+(n_1 n_2+n_3 n_4) \rho_{00} V+ (n_1 n_3+n_2 n_4) \rho_{00} 
S+ (n_1 n_4+n_2 n_3) \rho_{00}C\right](\frac{i\tau_2}{2})
\\
{\cal M}&= &(n_1+n_2+n_3+n_4)\frac{{\cal V}_6}{2}
\int_0^\infty \frac{d \tau_2}{\tau_2^4}\frac{1}{\eta^4}\left[
\frac{1}{2} \rho_{00}O+ \frac{1}{2}\rho_{01} +
\rho_{11}+\epsilon\rho_{10}\right]
(\frac{i\tau_2}{2}+\frac{1}{2}) \quad . \nonumber
\eea
Notice that for $\epsilon= +1$ terms proportional to $I_{V}^{2}$,
$I_{S}^{2}$, $I_{C}^{2}$ do not contribute to tadpoles and therefore
only the sum of
the CP charges is fixed: $I_{O}= n_1+n_2+n_3+n_4= 16$.  This
restricts only the total rank of the CP gauge group.
At the massless level one thus finds:\\ \\ 
\begin{tabular}{llll}
$\epsilon$&supersymmetry& Gauge group & Hypermultiplets \\
$+$ &{\cal N}= (1,1)& $\prod_{i= 1}^{4} Sp(n_{i})$ & $---$\\
$-$ &{\cal N}= (1,0)& $Sp(4)^4$ &
$({\bf 4},{\bf 4},{\bf 1},{\bf 1})+({\bf 4},{\bf 1},{\bf 4},{\bf 1})+
({\bf 4},{\bf 1},{\bf 1},{\bf 4})+$\\
 & & &$({\bf 1},{\bf 4},{\bf 4},{\bf 1})+({\bf 1},{\bf 4},{\bf 
1},{\bf 4}) 
+
 ({\bf 1},{\bf 1},{\bf 4},{\bf 4})$
\end{tabular}\\

The models with ${\cal N}= (1,1)$ correspond, at least topologically, 
to toroidal
compactifications without vector structure, \ie with a reduction of the
rank of the CP group associated to a non-vanishing generalized
second Stieffel-Whitney class and measured by the presence of a quantized
NS-NS antisymmetric tensor \cite{bpstor, mbtor}. 
The model with
${\cal N}= (1,0)$ is chiral
but anomaly-free thanks to the GSS mechanism \cite{gs,asano} that
involves several antisymmetric tensors and is the field-theory
counterpart of the R-R tadpole conditions \cite{mbas}.
Notice that because of the chiral $Z_{2}$ action, that implies a
quotienting by the T-duality transformation $X^{i}_{L}\rightarrow - 
X^{i}_{L}$ with
$X^{i}_{R}\rightarrow + X^{i}_{R}$ for $i=1\ldots 4$, would-be D9-
and D5-branes
are effectively identified and form some generalized brane bound-state.  
This
accounts for a further reduction by half of the rank of the CP group.  
As mentioned in the Introduction, it is a difficult task to
describe the above non-geometric vacuum configurations in terms of
D-branes and O-planes. Their concepts become fuzzy in highly curved or 
non-geometric backgrounds
\cite{fucsch, penta}. In the case under consideration, however, 
open-strings belonging to twisted sectors could be thought 
as strings with 
one end on a would-be D5-branes and the other end on the would-be 
D9-brane, here described by an unique CP charge $n_{i}$.
It is interesting to observe that the ${\cal N}= (1,1)$ vector 
multiplets are recovered in the $\epsilon=+1$ case 
by mixing the would-be D9-D9 and D5-D5
states with the would-be D9-D5 states. 

\section{$T^4/Z_3^L\times Z_3^R$ orbifold and
its open-string descendants}

Let us now consider the $T^4/Z_3^L\times Z_3^R$ orbifold of the type
IIB superstring. As a choice for a $T^{4}$ admitting a chiral $Z_3$
isometry we take the torus associated with the weight lattice of 
$SU(3)^{2}$.
This requires to turn on a quantized
NS-NS antisymmetric tensor background of rank four.
Notice that, differently from what happens for geometric orbifolds, we 
will
find the presence of open-string twisted sectors. Once again, no
compelling D-brane interpretation is available
for this kind of open strings in the present context.

The torus amplitude for this asymmetric orbifold is given by 
(\ref{torus6})
with
\bea
{\cal T}_{00}&= &\frac{1}{9}\left[ \rho_{00}\bar{\rho}_{00}
\Lambda_{SU(3)^2}+(\rho_{01}+\rho_{02})\bar{\rho}_{00}\bar{\Lambda}_{R}
+\rho_{00}(\bar{\rho}_{01}+\bar{\rho}_{02})\Lambda_{R}
+|\rho_{01}+\rho_{02}|^2\right]
\nonumber\\
{\cal T}_{01}&= &\frac{1}{9}\left[\rho_{00}
(\bar{\rho}_{10}\Lambda_{W}+\bar{\rho}_{11}\Lambda^\omega_{W}
+\bar{\rho}_{12}\Lambda^{\bar{\omega}}_{W})
+(\epsilon\rho_{01}+\bar{\epsilon}\rho_{02})(\bar{\rho}_{10}+
\bar{\rho}_{11}+\bar{\rho}_{12})\right]\nonumber\\
{\cal T}_{02}&= &\frac{1}{9}\left[\rho_{00}
(\bar{\rho}_{20}\Lambda_{W}+\bar{\rho}_{22}\Lambda^\omega_{W}
+\bar{\rho}_{21}\Lambda^{\bar{\omega}}_{W})
+(\epsilon\rho_{02}+\bar{\epsilon}\rho_{01})(\bar{\rho}_{20}+
\bar{\rho}_{22}+\bar{\rho}_{21})\right]\nonumber\\
{\cal T}_{11}&= &\frac{1}{9}\left[9(\rho_{10}\bar{\rho}_{10}
+\rho_{11}\bar{\rho}_{11}+\rho_{12}\bar{\rho}_{12})-3\epsilon
(\rho_{10}\bar{\rho}_{12}+\rho_{11}\bar{\rho}_{10}+
\rho_{12}\bar{\rho}_{11})\right.\nonumber\\
&&\left.-3\bar{\epsilon}
(\rho_{11}\bar{\rho}_{12}+\rho_{12}\bar{\rho}_{10}+
\rho_{10}\bar{\rho}_{11})\right]\nonumber\\
{\cal T}_{22}&= &\frac{1}{9}\left[9(\rho_{20}\bar{\rho}_{20}
+\rho_{22}\bar{\rho}_{22}+\rho_{21}\bar{\rho}_{21})-3{\epsilon}
(\rho_{20}\bar{\rho}_{21}+\rho_{22}\bar{\rho}_{20}+
\rho_{21}\bar{\rho}_{22})\right.\nonumber\\
&&\left.-3\bar{\epsilon}
(\rho_{22}\bar{\rho}_{21}+\rho_{21}\bar{\rho}_{20}+
\rho_{20}\bar{\rho}_{22})\right] \nonumber\\
{\cal T}_{12}&= &\frac{1}{9}\left[9(\rho_{10}\bar{\rho}_{20}
+\rho_{11}\bar{\rho}_{22}+\rho_{12}\bar{\rho}_{21})-3\epsilon
(\rho_{10}\bar{\rho}_{22}+\rho_{11}\bar{\rho}_{21}+
\rho_{12}\bar{\rho}_{20})\right.\nonumber\\
&&\left.-3\bar{\epsilon}
(\rho_{11}\bar{\rho}_{20}+\rho_{12}\bar{\rho}_{22}+
\rho_{10}\bar{\rho}_{21})\right] \quad .
\label{torus3}
\eea
The ${\cal T}_{10}$, ${\cal T}_{20}$ and ${\cal T}_{21}$ torus 
amplitudes
are given by the complex conjugate of ${\cal T}_{01}$, ${\cal T}_{02}$
and ${\cal T}_{12}$ respectively. 

As above, depending on the choice of $\epsilon$, the model enjoys 
${\cal
N}= (2,2)$ or ${\cal N}= (2,0)$ spacetime supersymmetry.
The resulting massless oriented closed-string contents are:\\

\begin{tabular}{lll }
$\epsilon$ & Supersymmetry & Supermultiplets \\
  $ 1$ & ${\cal N}= (2,2)$ & ${\bf G}_{(2,2)} $ \\
  $ e^{\pm{2\pi i \over 3}}$ & ${\cal N}= (2,0)$ & ${\bf G}_{(2,0)}+
  21\,{\bf T}_{(2,0)}$\\
  \end{tabular}\\
  
Notice that in this case the two choices 
$\epsilon=  e^{+{2\pi i \over 3}}$
and $ \epsilon=  e^{-{2\pi i \over 3}}$ give equivalent theories.
As before, the Klein-bottle amplitude is expressed only in terms of the 
chiral amplitudes $\delta_{g_L,g_R}
\rho_{g_L,h_L-h_R}$, that appear diagonally in (\ref{torus3}),
and reads
\bea
{\cal K}&= &\frac{{\cal V}_6}{2}
\int_0^\infty \frac{d \tau_2}{\tau_2^4}\frac{1}{\eta^4}\left[
\frac{1}{3}(\rho_{00}\Lambda_W+\rho_{01}+\rho_{02})\right.\nonumber\\
&&\left.+3\rho_{10}-\epsilon\rho_{11}
-\bar{\epsilon}\rho_{12}
+3\rho_{20}-\epsilon\rho_{22}-\bar{\epsilon}\rho_{21}
\right](2i\tau_2) \quad .
\label{klein3}
\eea
The action of $\Omega$ on the fixed points is determined by the
requirement that only $Z_3$ invariant states
flow in the transverse $\langle C|C\rangle$ amplitude. Indeed, writing 
(\ref{klein3})
in terms of the closed-string variables, one finds
\bea
{\widetilde{\cal K}}&= &2^3\frac{{\cal V}_6}{2}
\int_0^1 \frac{d q}{2\pi q}\frac{1}{\eta^4}\left[
\rho_{00}\Lambda_R+\rho_{01}+\rho_{02}\right.\nonumber\\
&&\left. +\rho_{10}+\bar{\epsilon}\rho_{11}+\epsilon\rho_{12}
+\rho_{20}+\epsilon\rho_{21}+\bar{\epsilon}\rho_{22}\right](q)
\label{kleint3} \quad .
\eea
Thus, only states with $Z_3$ eigenvalue equal to $1$ in the untwisted 
sector 
and $\epsilon, \bar{\epsilon}$ in the twisted sectors flow in the 
$\langle C|C\rangle$
amplitude.

The Klein-bottle projection (anti)symmetrizes
NS-NS (R-R) states
in the left-right symmetric 
sectors, ${\cal T}_{00}, {\cal T}_{11},{\cal 
T}_{22}$, and halves the ones in the remaining left-right asymmetric
sectors. The unoriented massless closed-string
states that survive the projections are given by:\\ \\
\begin{tabular}{lll }
$\epsilon$ & Supersymmetry & Supermultiplets \\
  $ 1$ & ${\cal N}= (1,1)$ & ${\bf G}_{(1,1)}+4\, {\bf 
V}^{c}_{(1,1)}$ \\
  $ e^{\pm{2\pi i \over 3}}$ & ${\cal N}= (1,0)$ & ${\bf 
G}_{(1,0)}+15\,
  {\bf H}_{(1,0)}+
6\,{\bf T}_{(1,0)}$\\
\end{tabular}\\ \\
According to our simplifying ansatz,
the open-string sectors are completely determined once (\ref{kleint3}) 
is given. The relevant
transverse amplitudes read
\bea
{\widetilde{\cal A}}&= &2^{-3}\frac{N^2}{2}{\cal V}_6
\int_0^1 \frac{d q}{2\pi q}\frac{1}{\eta^4}\left[
\rho_{00}\Lambda_R+\rho_{01}+\rho_{02}\right.\nonumber\\
&&\left. +\rho_{10}+\bar{\epsilon}\rho_{11}+\epsilon\rho_{12}
+\rho_{20}+\epsilon\rho_{21}+\bar{\epsilon}\rho_{22}\right](q)
\nonumber\\
{\widetilde{\cal M}}&= &-2 \frac{N}{2}{\cal V}_6
\int_0^1 \frac{d q}{2\pi q}\frac{1}{\eta^4}\left[
\rho_{00}\Lambda_R+\rho_{01}+\rho_{02}\right.\nonumber\\
&&\left. +\rho_{11}+\bar{\epsilon}\rho_{12}+\epsilon\rho_{10}
+\rho_{22}+\epsilon\rho_{20}+\bar{\epsilon}\rho_{21}\right](-q)
\quad ,
\eea
with $N= 8$. In the direct channel we are finally left with
\bea
{\cal A}&= &\frac{N^2}{2}{\cal V}_6
\int_0^\infty \frac{d \tau_2}{\tau_2^4}\frac{1}{\eta^4}
\left[\frac{1}{3}(\rho_{00}\Lambda_W+
\rho_{01}+\rho_{02})\right.\nonumber\\
&&\left.+3\rho_{10}-\epsilon \rho_{11}-\bar{\epsilon}\rho_{12}
+3\rho_{20}-\bar{\epsilon}\rho_{21}-\epsilon\rho_{22}
\right](\frac{i\tau_2}{2})
\label{annulus3} \nonumber\\
{\cal M}&= &-\frac{N}{2}{\cal V}_6
\int_0^\infty \frac{d \tau_2}{\tau_2^4}\frac{1}{\eta^4}
\left[\frac{1}{3}(\rho_{00}\Lambda_W+
\rho_{01}+\rho_{02})\right.\nonumber\\
&&\left. +3\rho_{11}-{\epsilon}\rho_{12}-\bar\epsilon\rho_{10}
+3\rho_{22}-\bar\epsilon\rho_{20}-{\epsilon}\rho_{21}
\right](\frac{i\tau_2}{2}+\frac{1}{2}) \quad .
\label{moebius3}
\eea
The additional phases are due to the fact that only ``hatted''
quantitities should enter ${\cal M}$ and
$\widetilde{\cal M}$ as required by reality of the amplitudes.

The resulting massless open-string content is now given by:\\ \\
\begin{tabular}{llll}
$\epsilon$&supersymmetry& Gauge group & Hypermultiplets \\
$1$ &(1,1)& $SO(8) $& $---$\\
$e^{\pm \frac{2 \pi i}{3}}$ &(1,0)&$SO(8)$ &  $4\,({\bf 28})$
\end{tabular}\\ \\
Once again, the model with ${\cal N}= (1,1)$ corresponds to a 
toroidal
compactification without vector structure \cite{bpstor,mbtor}, 
while the model with
${\cal N}= (1,0)$ is chiral
but anomaly-free thanks to the GSS mechanism \cite{gs,asano}.
Notice that because of the chiral $Z_{3}$ action
the open-string spectrum involves states belonging to the twisted
sectors. This should not sound 
too surprising given the form of the
torus amplitude. 

\section{$T^6/Z_3^L\times Z_3^R$ Orbifold and
its open-string descendants}

Very similarly to the above model, we can now discuss the
$Z_3^L\times Z_3^R$ orbifold of the type
IIB superstring in $D= 4$ and its open-string descendants. As a 
choice for
a $T^{6}$ admitting a chiral $Z_3$
isometry we take the torus associated with the weight lattice of 
$SU(3)^{3}$.
This requires to turn on a quantized
NS-NS antisymmetric tensor background of rank six.

The torus amplitude is given by
\be
{\cal T}= {\cal V}_4\int_{\cal F}\frac{d^2 \tau}{\tau_2^2}
X_4\sum_{g_L,g_R}
{\cal T}_{g_L,g_R} \quad ,
\label{torus4}
\ee
with
\bea
{\cal T}_{00}&= &\frac{1}{9}\left[ \rho_{00}\bar{\rho}_{00}
\Lambda_{SU(3)^2}+(\rho_{01}+\rho_{02})\bar{\rho}_{00}\bar{\Lambda}_{R}
+ \rho_{00}(\bar{\rho}_{01}+\bar{\rho}_{02})\Lambda_{R}+
|\rho_{01}+\rho_{02}|^2\right]
\nonumber\\
{\cal T}_{01}&= &\frac{1}{9}\left[\rho_{00}
(\bar{\rho}_{10}\Lambda_{W}+\bar{\rho}_{11}\Lambda^\omega_{W}
+\bar{\rho}_{12}\Lambda^{\bar{\omega}}_{W})
+(\epsilon\rho_{01}+\bar{\epsilon}\rho_{02})(\bar{\rho}_{10}+
\bar{\rho}_{11}+\bar{\rho}_{12})\right]\nonumber\\
{\cal T}_{02}&= &\frac{1}{9}\left[\rho_{00}
(\bar{\rho}_{20}\Lambda_{W}+\bar{\rho}_{22}\Lambda^\omega_{W}
+\bar{\rho}_{21}\Lambda^{\bar{\omega}}_{W})
+(\epsilon\rho_{02}+\bar{\epsilon}\rho_{01})(\bar{\rho}_{20}+
\bar{\rho}_{22}+\bar{\rho}_{21})\right]\nonumber\\
{\cal T}_{11}&= &\frac{1}{9}\left[27(\rho_{10}\bar{\rho}_{10}
+\rho_{11}\bar{\rho}_{11}+\rho_{12}\bar{\rho}_{12})+3\sqrt{3}i\epsilon
(\rho_{10}\bar{\rho}_{12}+\rho_{11}\bar{\rho}_{10}+
\rho_{12}\bar{\rho}_{11})\right.\nonumber\\
&&\left.-3\sqrt{3}i\bar{\epsilon}
(\rho_{11}\bar{\rho}_{12}+\rho_{12}\bar{\rho}_{10}+
\rho_{10}\bar{\rho}_{11})\right]\nonumber\\
{\cal T}_{22}&= &\frac{1}{9}\left[27(\rho_{20}\bar{\rho}_{20}
+\rho_{22}\bar{\rho}_{22}+\rho_{21}\bar{\rho}_{21})+3\sqrt{3}i\epsilon
(\rho_{20}\bar{\rho}_{21}+\rho_{22}\bar{\rho}_{20}+
\rho_{21}\bar{\rho}_{22})\right.\nonumber\\
&&\left.-3\sqrt{3}i\bar{\epsilon}
(\rho_{22}\bar{\rho}_{21}+\rho_{21}\bar{\rho}_{20}+
\rho_{20}\bar{\rho}_{22})\right]\nonumber\\
{\cal T}_{12}&= &\frac{1}{9}\left[27(\rho_{10}\bar{\rho}_{20}
+\rho_{11}\bar{\rho}_{22}+\rho_{12}\bar{\rho}_{21})-3\sqrt{3}i\epsilon
(\rho_{10}\bar{\rho}_{22}+\rho_{11}\bar{\rho}_{21}+
\rho_{12}\bar{\rho}_{20})\right.\nonumber\\
&&\left.+3\sqrt{3}i\bar{\epsilon}
(\rho_{11}\bar{\rho}_{20}+\rho_{12}\bar{\rho}_{22}+
\rho_{10}\bar{\rho}_{21})\right] \quad .
\eea
The amplitudes ${\cal T}_{10}$, ${\cal T}_{20}$ and ${\cal T}_{21}$ 
are the complex conjugate of ${\cal T}_{01}$, ${\cal T}_{02}$
and ${\cal T}_{12}$, respectively.

Depending on the choice of discrete torsion $\epsilon$,
the massless oriented closed-string spectra are given by:\\ \\
\begin{tabular}{lll }
$\epsilon$ & Supersymmetry & Supermultiplets \\
$1$ & ${\cal N}= 4$ & ${\bf G}_{4}+10 {\bf V}_{4}$ \\
$e^{+\frac{2\pi i}{3}}$ &
${\cal N}= 2$ & ${\bf G}_{2}+19\,{\bf H}_{2}+6\,{\bf V}_{2}$\\
$e^{-\frac{2\pi i}{3}}$ &
${\cal N}= 2$ & ${\bf G}_{2}+7\,{\bf H}_{2}+18\,{\bf V}_{2}$\\
\end{tabular}\\ \\
and correspond,
respectively, to compactifications on spaces topologically
equivalent to $K3\times T^{2}$
and to two mirror Calabi-Yau spaces with Hodge numbers
$h_{1,1}= 18=  h_{1,2}^{\prime}$, $h_{1,2}= 6= h_{1,1}^{\prime}$.

Notice that, differently from \cite{dabhar}, we find for $\epsilon= 1$ 
an
 ${\cal N}= 4$ theory with only 10 rather than 28 vector multiplets.
 The number of vector multiplets is smaller than the maximum number, 
22,
 found in superstring compactifications with ${\cal N}= 4$ in 
$D= 4$.
From the type II perspective, they correspond to compactifications
on manifolds that are locally, but not necessarily globally, of the
form $K3\times T^{2}$. From the type I or, equivalently, heterotic
perspective, they correspond to toroidal compactifications possibly 
with
 non-commuting Wilson lines \cite{bpstor, mbtor, nsw,  chl}.
 In fact, some candidate dual pairs descend from the duality 
 between type IIA superstring on $K3$ and heterotic string on $T^{4}$
 \cite{chalow}. In the case under
 consideration, because of the chiral nature of the projections and
 twistings, it is not easy to find a geometric interpretation for the
 above superstring vacuum configurations. Still, the left-right
 symmetry of the resulting oriented closed-string theory suggests that
 there is no reason to exceed the ``experimental'' bound of 22 on the 
number of vector multiplets. Moreover no ``exotic'' brane whose
 excitations could account for the extra vector multiplets has been 
 proposed so far \cite{dabhar}.

Let us therefore discuss the open-string descendants.
The Klein-bottle projection is now given by
\bea
{\cal K}&= &\frac{{\cal V}_4}{2}
\int_0^\infty \frac{d \tau_2}{\tau_2^4}\frac{1}{\eta^4}\left[
\frac{1}{3}(\rho_{00}\Lambda_W+\rho_{01}+\rho_{02})\right.\nonumber\\
&&\left.+9\rho_{10}+i\sqrt{3}\epsilon\rho_{11}-
i\sqrt{3}\bar{\epsilon}\rho_{12}
+9\rho_{20}+i\sqrt{3}\epsilon\rho_{22}-i\sqrt{3}\bar{\epsilon}\rho_{21}
\right](2i\tau_2) \quad ,
\eea
and yields the following unoriented massless closed-string spectra: \\ \\
\begin{tabular}{lll }
$\epsilon$ & Supersymmetry & Supermultiplets \\
$1$ & ${\cal N}= 2$ & ${\bf G}_{2}+2 {\bf V}_{2}+9 {\bf H}_2$ \\
$e^{+\frac{2\pi i}{3}}$ &
${\cal N}= 1$ & ${\bf G}_{1}+25\,{\bf C}_{1}$\\
$e^{-{2\pi i \over 3}}$ &
${\cal N}= 1$ & ${\bf G}_{1}+22\,{\bf C}_{1}+3\,{\bf V}_{1}$\\
\end{tabular}\\ \\
with ${\bf C}_{1}$ denoting the chiral multiplet in $D= 4$.

The type I descendant is constructed by identifying the
transverse-channel amplitudes
\bea
{\widetilde{\cal K}}&= &2^2\frac{\sqrt{3}}{2}{\cal V}_4
\int_0^1 \frac{d q}{2\pi q}\frac{1}{\eta^2}\left[
\rho_{00}\Lambda_R+\rho_{01}+\rho_{02}\right.\nonumber\\
&&\left. +\rho_{10}+\bar{\epsilon}\rho_{11}+\epsilon\rho_{12}
+\rho_{20}+\epsilon\rho_{21}+\bar{\epsilon}\rho_{22}\right](q)
\nonumber\\
{\widetilde{\cal A}}&= &2^{-2}\sqrt{3}\frac{N^2}{2}{\cal V}_4
\int_0^1 \frac{d q}{2\pi q}\frac{1}{\eta^2}\left[
\rho_{00}\Lambda_R+\rho_{01}+\rho_{02}\right.\nonumber\\
&&\left. +\rho_{10}+\bar{\epsilon}\rho_{11}+\epsilon\rho_{12}
+\rho_{20}+\epsilon\rho_{21}+\bar{\epsilon}\rho_{22}\right](q)
\nonumber\\
{\widetilde{\cal M}}&= &-2\sqrt{3}\frac{N}{2}{\cal V}_4
\int_0^1 \frac{d q}{2\pi q}\frac{1}{\eta^2}\left[
\rho_{00}\Lambda_R+\rho_{01}+\rho_{02}\right.\nonumber\\
&&\left. +\rho_{11}+\bar{\epsilon}\rho_{12}+\epsilon\rho_{10}
+\rho_{22}+\epsilon\rho_{20}+\bar{\epsilon}\rho_{21}\right](-q)
\quad ,
\eea
with one another. This requires taking $N= 4$.

The direct-channel open-string amplitudes then read
\bea
{\cal A}&= &\frac{N^2}{2}{\cal V}_4
\int_0^\infty \frac{d \tau_2}{\tau_2^4}\frac{1}{\eta^2}
\left[\frac{1}{3}(\rho_{00}\Lambda_W
+\rho_{01}+\rho_{02})\right.\nonumber\\
&&\left.+9\rho_{10}+i\sqrt{3}\epsilon 
\rho_{11}-i\sqrt{3}\bar{\epsilon}
\rho_{12}
+9\rho_{20}-i\sqrt{3}\bar{\epsilon}\rho_{21}+i\sqrt{3}\epsilon\rho_{22}
\right](\frac{i\tau_2}{2})
\nonumber\\
{\cal M}&= &+\frac{N}{2}{\cal V}_4
\int_0^\infty \frac{d \tau_2}{\tau_2^4}\frac{1}{\eta^2}
\left[\frac{1}{3}(\rho_{00}\Lambda_W
+\rho_{01}+\rho_{02})\right.\nonumber\\
&&\left.+9\rho_{11}+i\sqrt{3}\epsilon 
\rho_{12}-i\sqrt{3}\bar{\epsilon}
\rho_{10}
+9\rho_{22}-i\sqrt{3}\bar{\epsilon}\rho_{20}+i\sqrt{3}\epsilon\rho_{21}
\right](\frac{i\tau_2}{2}+\frac{1}{2}) \quad .
\label{moebius43}
\eea
The resulting CP group\footnote{This corrects an imprecise
statement made in the last section of \cite{chiral} 
concerning $Z_{3}$ orbifolds with quantized $B_{ij}$.} is $Sp(4)$ and the massless open-string content is
given by\\

\begin{tabular}{llll}
$\epsilon$ & Supersymmetry &Gauge Group& Hypermultiplets\\
$1$ & ${\cal N}= 2$ & $Sp(4)$&$ 4({\bf 10})$ \\
$e^{+\frac{2\pi i}{3}}$ &
${\cal N}= 1$ & $ Sp(4)$ & $6 ({\bf 10})$\\
$e^{-\frac{2\pi i}{3}}$ &
${\cal N}= 1$ & $Sp(4)$&$ 12({\bf 10})$\\
\end{tabular}

The $\epsilon= 1$ model is IR free. The CP group can be completely 
higgsed in
the Higgs branch and the remaining massless hypermultiplets
parametrize a non-trivial quaternionic manifold which becomes
hyperk\"aler in the rigid limit.
The models with $\epsilon=  e^{\pm{2\pi i\over3}}$ are non chiral and
IR free. Although phenomenologically not very appealing, they are
interesting in at least two respects. First, states belonging to the
twisted sectors of the chiral orbifold group appear in the open-string
spectrum. Second, at least for $\epsilon=  e^{-{2\pi i\over3}}$,
additional vector multiplets appear in the unoriented closed-string
spectrum thus decreasing the number of marginal deformations.
The additional vectors belong to the R-R sector and as such are not
minimally coupled to the perturbative string excitations.
It would be interesting to identify the non-perturbative states
that couple minimally to these R-R fields. They would represent
the generalization of the concept of D-brane in these non-geometric 
compactifications.

\section{$T^{4}/Z^L_2\times Z_2^R$ with non-supersymmetric 
open-string 
sectors}

In recent times, there has been a lot of interest in constructing
type I models with brane supersymmetry breaking either at the 
string scale \cite{bsb} or at the compactification scale
\cite{bach, penta, iban}. 
By this one means models in which supersymmetry is exact in the bulk but it is
broken along (a subset of) the branes\footnote{Alternatively one may 
consider models in which the theory in the bulk is less supersymmetric than 
the theory on the branes \cite{ads}.}. 
Clearly, because of the
coupling between the brane excitations and the bulk modes, 
supersymmetry breaking is then transmitted to the whole system.
So far, only geometric models have been considered.
We would like to observe that the same mechanism of supersymmetry breaking can
be exposed in the context of non-geometric models such as the ones 
discussed in the present paper.
To keep the discussion as simple as possible we will restrict our 
attention to a non-supersymmetric version of the open-string descendant
of the $T^{4}/ Z_2^L\times Z_2^R$ orbifold
 discussed in Section 2. Supersymmetry breaking can be
achieved by replacing the ``$D$-branes'', associated
to the $n_3,n_4$ CP charges, with ``anti-D-branes'', and
modifying the $\Omega$ action on the fixed-point space. The
resulting unoriented descendant is encoded in the following amplitudes
\bea
{\cal K}_{ns}&= &\frac{{\cal V}_6}{2}
\int_0^\infty \frac{d \tau_2}{\tau_2^4}\frac{1}{\eta^4}\left[
\frac{1}{2}\rho_{00}(O+V-S-C)+\frac{1}{2}\rho_{01}
+2\epsilon\rho_{11}\right](2i\tau_2)
\nonumber\\
{\cal A}_{ns}&= &\frac{{\cal V}_6}{2}
\int_0^\infty \frac{d \tau_2}{\tau_2^4}\frac{1}{\eta^4}\left[
(2n\bar{n}+2m\bar{m})(\frac{1}{2}\rho_{00} O+ \frac{1}{2}\rho_{01}
+\rho_{10}+\epsilon\rho_{11})\right.\nonumber\\
&&\left.
+(n^2+\bar{n}^2+m^2+\bar{m}^2)(\frac{1}{2}\rho_{00} V+
\rho_{10}-\epsilon\rho_{11})
\right.\nonumber\\
&&\left.+ (2 n \bar{m}+2\bar{n}m)(\frac{1}{2}\rho^{\prime}_{00} S +
\rho^{\prime}_{10}
-\epsilon\rho^{\prime}_{11})
+(2 n m+ 2\bar{n}\bar{m})(\frac{1}{2}\rho^{\prime}_{00} C +
\rho^{\prime}_{10}
-\epsilon\rho^{\prime}_{11})
\right](\frac{i\tau_2}{2})
\nonumber\\
{\cal M}_{ns}&= &\frac{{\cal V}_6}{2}
\int_0^\infty \frac{d \tau_2}{\tau_2^4}\frac{1}{\eta^4}\left[
(n+\bar{n})(- \frac{1}{2}\rho_{00}V+\rho_{11}-\epsilon\rho_{10})
\right.\nonumber\\
&&\left.
-(m+\bar{m})(\frac{1}{2}{\rho}^{\prime\prime}_{00}V+
{\rho}^{\prime\prime}_{11}-\epsilon
{\rho}^{\prime\prime}_{10})\right]
(\frac{i\tau_2}{2}+\frac{1}{2}) \quad ,
\eea
where with a prime ($^{\prime}$) and a double prime 
($^{\prime\prime}$)
we denote non-supersymmetric chiral traces 
obtained by modifying the sums over spin structures in $\rho_{g,h}$
in the way discussed in Appendix A.  This correspond to changing the sign of 
the Ramond sector for the
chiral supertraces with a double prime and to interchanging $O,S$ with 
$V,C$ 
in the space-time part for the
amplitudes with a prime. 
The transverse amplitudes are then given by
\bea
{\widetilde{\cal K}}_{ns}&= &2^3\frac{{\cal V}_6}{2}
\int_0^1 \frac{d q}{2\pi q}\frac{1}{\eta^4}\left[
\rho_{00}V+2\rho_{10}-2\epsilon \rho_{11}\right](q)
\nonumber\\
{\widetilde{\cal A}}_{ns}&= &2^{-3}\frac{{\cal V}_6}{2}
\int_0^1 \frac{d q}{2\pi q}\frac{1}{\eta^4}\left[
\frac{1}{8}(I_{O}^2+I_{V}^2)(\rho_{00}(O+V)+\rho_{01}+4\rho_{10})
\right.\nonumber\\&&\left.
+\frac{1}{8}(I_{O}^2-I_{V}^2)({\rho}^{\prime\prime}_{00}(O-V)+
{\rho}^{\prime\prime}_{01}+
4\epsilon{\rho}^{\prime\prime}_{11})
\right.\nonumber\\
&&\left.-\frac{1}{8}(I_{S}^2+I_{C}^2)(\rho_{00}(S+C)+4\rho_{10}-4\epsilon
\rho_{11})
-\frac{1}{8}(I_{S}^2-I_{C}^2){\rho}^{\prime\prime}_{00}(S-C)\right](q)
\nonumber\\
{\widetilde{\cal M}}_{ns}&= &-2\frac{{\cal V}_6}{2}
\int_0^1 \frac{d q}{2\pi q}\frac{1}{\eta^4}\left[
\frac{(I_{V}+I_{O})}{4}(-\rho_{00}V+2\rho_{11}-2\epsilon\rho_{10})
\right.\nonumber\\&&\left.
+\frac{(I_{V}-I_{O})}{4}(\rho_{00}^{\prime\prime}V+2
\rho_{11}^{\prime\prime}
-2\epsilon\rho_{10}^{\prime\prime})
\right](-q) \quad ,
\eea 
where it is convenient to parametrize the CP charge assignments as  
\bea
I_{0}&= &n+\bar{n}+m+\bar{m}\nonumber\\
I_{V}&= &n+\bar{n}-m-\bar{m}\nonumber\\
I_{S}&= &n-\bar{n}+m-\bar{m}\nonumber\\
I_{C}&= &n+\bar{n}-m+\bar{m} \quad ,
\eea
with $n= \bar{n} = m= \bar{m}= 4$ fixed by R-R tadpole 
cancellation.
Comparing the above amplitudes in the transverse channel with the
ones found in the supersymmetric case
one can see that the replacements correspond precisely to flipping the
signs of the mixed $(nm)$ RR-sectors, so that
$\rho_{gh}\rightarrow{\rho}^{\prime\prime}_{gh}$.  As already
discussed, this is interpreted as
a replacement of a (sub)set of  would-be D-branes by
the corresponding  would-be anti-D-branes.
Moreover, as mentioned in the introduction, we have relaxed our
simplifying ansatz
${\widetilde{\cal K}}= {\widetilde{\cal A}}= {\widetilde{\cal M}}$.
Imposing this condition would have resulted into the supersymmetric model of
Section 2 and would not have reached our aim of
implementing the mechanism of brane supersymmetry breaking.

Depending on the choice of discrete torsion, the new Klein-bottle 
projection leads to the following massless unoriented
closed-string spectra: \\ 

\begin{tabular}{lll }
$\epsilon$ & Supersymmetry & Supermultiplets \\
  $ +$ & ${\cal N}= (1,1)$ & ${\bf G}_{(1,1)}+4\, {\bf 
V}^{c}_{(1,1)}$ \\
  $ -$ & ${\cal N}= (1,0)$ & ${\bf G}_{(1,0)}+10\,{\bf 
H}_{(1,0)}+11\,
{\bf T}_{(1,0)}$\\
\end{tabular}\\

Once again, let us stress that supersymmetry is exact in the bulk 
while,
as expected, it is broken in the open-string sector.
At the massless level one finds \\
{\bf $\epsilon= 1$} \\
\begin{tabular}{ll}  
matter & $U(4)^2$-representations \\
(V+4O-2S-2C) & $({\bf 16},{\bf 1})+({\bf 1},{\bf 16})$\\
2 (4O)& $ ({\bf 4},{\bf 4}) +  (\bar{\bf 4},\bar{\bf 4})+
({\bf 4},\bar{\bf 4}) +  (\bar{\bf 4},{\bf 4}) $\\  
\end{tabular}\\
\\
{\bf $\epsilon= -1$} \\
\begin{tabular}{ll} 
matter & $U(4)^2$-representations \\
(V-2S) & $({\bf 16},{\bf 1})+({\bf 1},{\bf 16})$\\
(4O-2C)& $({\bf 10},{\bf 1}) + (\bar{\bf 10},{\bf 1}) $\\  
(4O-2S)& $ ({\bf 4},{\bf 4}) +  (\bar{\bf 4},\bar{\bf 4})+
 ({\bf 4},\bar{\bf 4}) +  (\bar{\bf 4},{\bf 4}) $\\
(4O)&$({\bf 1},{\bf 6}) + ({\bf 1},\bar{\bf 6})$\\
(-2C)&$({\bf 1},{\bf 10}) + ({\bf 1},\bar{\bf 10})$
\end{tabular}\\

Notice, in particular, that supersymmetry is broken only by open strings
charged under the anti D-brane gauge group (the second $U(4)$ factor 
above). 
It is easy to check that both gauge and gravitational anomalies are
absent, thanks to the vanishing of the R-R tadpoles.
Indeed, for the potential
gauge anomaly of each $U(4)$ CP group one finds $2n-2(n+8)+4m= 0$,
while for the potential
gravitational anomaly, one finds 
$(4\times 10-4\times 16-2\times 16)_o+(29\times 11+10-273)_c= 0$.
Once the minimal ansatz is left aside, it is interesting to exploit 
some 
extra freedom available in the $\epsilon= 1$ case. Since, as one can 
immediately see, no massless
tadpoles are present in the transverse-channel Klein-bottle amplitude,
there is no need of adding open strings and no brane 
supersymmetry breaking is induced in this case.
The resulting six dimensional model is one of the most economic
string realizations of $N= (1,1)$ supergravity coupled to
4 vector supermultiplets.

Although there is no claim of phenomenological appeal,
we would like to stress that the potential applications of the above
construction to non-geometric models are undoubtly far reaching
and deserve an extensive study. It would be interesting to study
the connection of the above kind of models with stable non-BPS
configurations of branes studied in the recent literature (see \eg
\cite{sen} and references therein).

\section{Conclusions and discussion}

We have discussed some asymmetric orbifolds of the type IIB
superstring and their open and unoriented string descendants.

A nice feature of the models is that the introduction of discrete
torsion allows to break half of the supersymmetries and relate
``topologically'' different compactifications.
We have also shown that prior to the introduction of a non-trivial
discrete torsion no exotic phenomena \cite{dabhar} appear in the oriented
closed-string theories. This has been obtained after carefully taking
into account the subtle phases that appear in the modular
transformations of the chiral amplitudes.

From the algebraic point of view, the introduction of discrete
torsion allows one to relate
modular combinations of characters with extended symmetry and
permutation modular invariants. This gives a rationale for the
various unoriented descendants found by varying the choice of the
modular invariant one-loop amplitudes in the type II parent
theories.
In particular, one is lead to speculate that some of the permutation
modular invariants can be found by chiral projections much in the same
way as in the simple asymmetric orbifold instances that have been
considered above. In these cases the perturbatively different
unoriented descendants may turn out to be non-perturbatively
equivalent when a proper action of the projection is implemented on
the solitonic spectra and interactions.

The consistency
of the construction does not require a D-brane interpretation,
that
would be fuzzy in non-geometric environments such as the above ones
and to some extent
useless. At the string level, \ie when high curvatures are present or
when the vacuum configuration has no direct geometric interpretation,
algebraic constructions such as boundary and crosscap states
that can be extracted via techniques of unoriented descendants are much
more rewarding. It is worth stressing that, although the geometrical 
meaning of boundaries and crosscaps in the present context is obscure, 
string techniques \cite{mss} can still allow one to determine the
WZ anomalous couplings of the properly generalized solitonic objects 
to the bulk fields.
The question of identifying the non-perturbative (non)BPS solitons
in the above non-geometric backgrounds is still open.

\section{Acknowledgements}

We would like to thank 
C.~Angelantonj, A.~Dabholkar, A.~Hammou, K.S.~Narain, K.~Ray, A.~Sagnotti,
H.~Sarmadi, C.A.~Scrucca, M.~Serone and Ya.S.~Stanev for fruitful 
discussions.
We would like to thank the organizers of the ``Extended Workshop
in String Theory'' for the kind hospitality at Abdus Salam ICTP,
Trieste, where this work has been initiated.  One of us (M.B.) 
would like to thank the organizers of the workshop on
``String Theory: Duality, Gravity \& Field Theory'' 
for the stimulating atmosphere at the Aspen Center for Physics, where part 
of this investigation was carried on. 

\section{Appendix A: Conformal blocks in $D= 4,6$}

In a $Z_{N}$-orbifold, the chiral traces of an element $h$ over states
in a given $g$-twisted sector read 
\bea
\rho_{00}&\equiv& \frac{1}{2}\sum_{\alpha,\beta= 0,1/2}
(-)^{2\alpha+2\beta+4\alpha\beta}
\frac{\vartheta{\alpha \brack \beta}^4}{\eta^4}\nonumber\\
\rho_{0h}&\equiv&\frac{1}{2}\sum_{\alpha,\beta= 0,1/2}
(-)^{2\alpha+2\beta+4\alpha\beta}
\left(\frac{\vartheta{\alpha \brack \beta}}{\eta}\right)^{4-{d/2}}
 \,\prod_{i= 1}^{d/2} (2sin\pi h_i)\frac{\vartheta{\alpha \brack 
\beta+h_i}}
{\vartheta{\frac{1}{2} \brack \frac{1}{2}+h_i}}~~~h\neq 0\nonumber\\
\rho_{gh}&\equiv&-(i)^{\frac{d}{2}}\frac{1}{2}\sum_{\alpha,\beta= 0,1/2}
(-)^{2\alpha+2\beta+4\alpha\beta}
\left(\frac{\vartheta{\alpha \brack \beta}}{\eta}\right)^{4-{d/2}}
\, \prod_{i= 1}^{d/2} \frac{\vartheta{\alpha+g_i \brack \beta+h_i}}
{\vartheta{\frac{1}{2}+g_i \brack \frac{1}{2}+h_i}}
~~~g,h\neq 0\label{rho} \quad ,
\eea
where ${\vartheta{\alpha \brack \beta}}$ are the standard Jacobi
theta functions with characteristics and
 $\sum_{i}^{d/2} g_i=  \sum_{i}^{d/2}h_i = 0 (mod 1)$.

The behaviour under S-modular transformations ($\tau\rightarrow
-{1}/{\tau}$) is as follows
\bea
\rho_{00} &\rightarrow& \rho_{00}\nonumber\\
\rho_{0h} &\rightarrow
& (2 sin{\pi h})^{\frac{d}{2}}\, \rho_{h0}~~~h\neq 0\nonumber\\
\rho_{ho} &\rightarrow
& (2 sin{\pi h})^{-\frac{d}{2}}\, \rho_{0,-h}~~~h\neq 0\nonumber\\
\rho_{gg} &\rightarrow&
(i)^{\frac{d}{2}}\rho_{g,-g}~~~g\neq 0\nonumber\\
\rho_{g,-g} &\rightarrow
& (-i)^{\frac{d}{2}}\rho_{-g,-g}~~~g\neq 0 \quad .
\eea
The behaviour under T-modular transformations ($\tau\rightarrow
\tau + 1$) is as follows
\be
\eta^{-{D-2\over 2}} \rho_{gh}  \rightarrow
\eta^{-{D-2\over 2}} \rho_{g,g+h} \quad .
\ee
The modular transformation ${P}= ST^2ST$ then relates
chiral traces in the transverse and direct M\"obius-strip amplitudes
and corresponds to $\widehat{P}= T^{1/2}ST^2ST^{1/2}$
on ``hatted'' real characters \cite{mbas}.
 
The chiral traces entering the non-supersymmetric models (Section 6) 
are defined by
\bea
\rho^{\prime}_{gh} &\equiv& 
\frac{1}{2}\sum_{\alpha,\beta}
(-)^{2\alpha+4\alpha\beta}
\frac{\vartheta{\alpha \brack \beta}^{2}}{\eta^{2}}\prod_{i= 1}^2
\frac{\vartheta{\alpha+g_i \brack \beta+h_i}}
{\vartheta{\frac{1}{2}+g_i \brack \frac{1}{2}+h_i}}~~~g,h\neq 0\nonumber\\
{\rho}^{\prime\prime}_{gh} &\equiv& 
\frac{1}{2}\sum_{\alpha=
,\beta}
(-)^{2\beta+4\alpha\beta}
\frac{\vartheta{\alpha \brack \beta}^{2}}{\eta^{2}}\prod_{i= 1}^2
\frac{\vartheta{\alpha+g_i \brack \beta+h_i}}
{\vartheta{\frac{1}{2}+g_i \brack \frac{1}{2}+h_i}}~~~g,h\neq 0 
\quad ,
\eea  
with similar replacements for the remaining traces with $g$ and/or $h$ 
equal to zero.

Some relevant lattice sums for compact $SO(8)$ bosons are
\bea
\Lambda_{SO(8)}&= &|O_8|^2+|V_8|^2+|S_8|^2+|C_8|^2\nonumber\\
\Lambda_{W}^{\pm}&= &O_8\pm V_8\pm S_8\pm C_8
\quad ,
\eea
where $O_{n}, V_{n}, S_{n}, C_{n}$ are $SO(n)$ characters at level 
one. Some relevant lattice sums for compact $SU(3)^{\ell}$ bosons are
\bea
\Lambda_{SU(3)^{\ell}}&= &(|\chi_1|^2+|\chi_3|^2+
|\chi_{\bar{3}}|^2)^{\ell}
\nonumber\\
\Lambda_{R}&= &\chi_1^{\ell}\nonumber\\
\Lambda^\omega_{W}&= &(\chi_1+\omega\chi_3+
\omega\chi_{\bar{3}})^{\ell} \quad ,
\eea
where $\chi_{\bf 1}, \chi_{\bf 3}, \chi_{\bf \bar{3}}$ are
$SU(3)$ characters at level one.

\section{Appendix B: Open-string descendant of $T^4/Z_2^L\times 
Z_2^R$}

The $T^4/Z_2^L\times Z_2^R$ model with
$SO(8)$ lattice is a T-duality orbifold of the geometric $Z_{2}$
orbifold.
Denoting by $Q_{O}, Q_{V}, Q_{S}, Q_{C}$ the supersymmetric 
characters
\bea
Q_{O} &= & V_{4}O_{4} - C_{4}C_{4} \nonumber \\
Q_{V} &= & O_{4}V_{4} - S_{4}S_{4} \nonumber \\
Q_{S} &= & O_{4}C_{4} - S_{4}O_{4} \nonumber \\
Q_{C} &= & V_{4}S_{4} - C_{4}V_{4} \quad ,
\eea
one is lead to introduce the following 16 characters \cite{mbas}
\bea
\chi_{1} &= & Q_{O} O_{4}O_{4} + Q_{V} V_{4}V_{4} \qquad \qquad
\widetilde\chi_{1} =  Q_{S} S_{4}O_{4} + Q_{C} C_{4}V_{4} \nonumber 
\\
\chi_{2} &= & Q_{O} O_{4}V_{4} + Q_{V} V_{4}O_{4} \qquad \qquad
\widetilde\chi_{2} =  Q_{S} S_{4}V_{4} + Q_{C} C_{4}O_{4} \nonumber 
\\
\chi_{3} &= & Q_{O} C_{4}C_{4} + Q_{V} S_{4}S_{4} \qquad \qquad
\widetilde\chi_{3} =  Q_{S} V_{4}C_{4} + Q_{C} O_{4}S_{4} \nonumber 
\\
\chi_{4} &= & Q_{O} C_{4}S_{4} + Q_{V} S_{4}C_{4} \qquad \qquad
\widetilde\chi_{4} =  Q_{S} V_{4}S_{4} + Q_{C} O_{4}C_{4} \nonumber 
\\
\chi_{5} &= & Q_{O} V_{4}V_{4} + Q_{V} O_{4}O_{4} \qquad \qquad
\widetilde\chi_{5} =  Q_{S} C_{4}V_{4} + Q_{C} S_{4}O_{4} \nonumber 
\\
\chi_{6} &= & Q_{O} V_{4}O_{4} + Q_{V} O_{4}V_{4} \qquad \qquad
\widetilde\chi_{6} =  Q_{S} C_{4}O_{4} + Q_{C} S_{4}V_{4} \nonumber 
\\
\chi_{7} &= & Q_{O} S_{4}S_{4} + Q_{V} C_{4}C_{4} \qquad \qquad
\widetilde\chi_{7} =  Q_{S} O_{4}S_{4} + Q_{C} V_{4}C_{4} \nonumber 
\\
\chi_{8} &= & Q_{O} S_{4}C_{4} + Q_{V} C_{4}S_{4} \qquad \qquad
\widetilde\chi_{8} =  Q_{S} O_{4}C_{4} + Q_{C} V_{4}S_{4} \quad .
\eea
The chiral $Z_{2}$ generators act by $Q_{V}\rightarrow
-Q_{V}$, $Q_{C}\rightarrow -Q_{C}$, on the spacetime characters and
by $V_{4}\rightarrow -V_{4}$, $C_{4}\rightarrow -C_{4}$ on the second
$SO(4)$ factor in the decomposition of the internal $SO(8)$ into 
$SO(4)^{2}$. In the character basis the $Z_{2}$ generators
then act diagonally with plus eigenvalues for 
$\chi_{i},\tilde{\chi}_{i}$, with $i= 1,4,6,7$ and minus eigenvalues for the
remaining ones.   
The two modular invariant combinations, corresponding to the presence
or absence of discrete torsion between the two chiral
$Z_{2}$'s, can be defined by projecting onto states with
$Z_{2}^{L}= \epsilon Z_{2}^{R}= 1$ ($Z_{2}^{L}= \epsilon 
Z_{2}^{R}= -1$)
in the (un)twisted sector.
For {\bf $\epsilon= +1$}, one finds
\bea
{\cal T}&= &
{\cal V}_6\int_{\cal F}\frac{d^2 \tau}{\tau_2^2}\,X_6 \, [
\,|\chi_1|^2+|\chi_4|^2+|\chi_6|^2+|\chi_7|^2+|\tilde{\chi}_2|^2
+|\tilde{\chi}_3|^2+|\tilde{\chi}_5|^2+|\tilde{\chi}_8|^2\nonumber\\
&&+\chi_1\bar{\tilde{\chi}}_8+\chi_4\bar{\tilde{\chi}}_5+
\chi_6\bar{\tilde{\chi}}_3
+\chi_7\bar{\tilde{\chi}}_2+
+\tilde{\chi}_8\bar{\chi}_1+\tilde{\chi}_5\bar{\chi}_4+
\tilde{\chi}_3\bar{\chi}_6
+\tilde{\chi}_2\bar{\chi}_7 \, ] \quad ,
\label{e1}
\eea
while, for {\bf $\epsilon= -1$}, one finds
\bea
{\cal 
T}&= &
{\cal V}_6\int_{\cal F}\frac{d^2 \tau}{\tau_2^2}\,X_6 \,
[ \, |\chi_1|^2+|\chi_4|^2+|\chi_6|^2+|\chi_7|^2+|\tilde{\chi}_1|^2
+|\tilde{\chi}_4|^2+|\tilde{\chi}_6|^2+|\tilde{\chi}_7|^2\nonumber\\
&&+\chi_2\bar{\tilde{\chi}}_3+\chi_3\bar{\tilde{\chi}}_2+
\chi_5\bar{\tilde{\chi}}_8
+\chi_8\bar{\tilde{\chi}}_5+
+\tilde{\chi}_2\bar{\chi}_3+\tilde{\chi}_3\bar{\chi}_2+
\tilde{\chi}_5\bar{\chi}_8
+\tilde{\chi}_8\bar{\chi}_5 \, ] \quad .
\label{e-1}
\eea
One can check that the $\epsilon= -1$ combination corresponds to the
permutation modular invariant denoted by $A_{16}$ in \cite{gepner}.

Notice that for $\epsilon= 1$ the torus amplitude
can be rewritten as
\be
{\cal T}= {\cal V}_6\int_{\cal F}\frac{d^2 \tau}{\tau_2^2}\,X_6
\, [ \, |\xi_O|^2+|\xi_V|^2+ |\xi_S|^2+|\xi_C|^2 \, ]
\ee
in terms of the extended characters
\bea
\xi_O&= &\chi_1+\tilde{\chi}_8 =   Q O_{8}\nonumber\\
\xi_V&= &\chi_6+\tilde{\chi}_3 =   Q V_{8}\nonumber\\
\xi_S&= &\chi_7+\tilde{\chi}_2 =   Q S_{8}\nonumber\\
\xi_C&= &\chi_4+\tilde{\chi}_5 =   Q C_{8} \quad ,
\eea
where $Q= V_{8}-S_{8}$, reflecting the fact that the orbifold 
correspond to a toroidal compactification of the type IIB superstring.

For $\epsilon= 1$, the corresponding open-string descendant (for the 
simplest CP group assignments) is then given by\\
\bea
{\cal K}&= &\frac{{\cal V}_6}{2}
\int_0^\infty \frac{d \tau_2}{\tau_2^4}\frac{1}{\eta^4}\left[
\xi_O+\xi_V+\xi_S+\xi_C\right](2i\tau_2)\nonumber\\
{\cal A}&= &\frac{{\cal V}_6}{2}
\int_0^\infty \frac{d \tau_2}{\tau_2^4}\frac{1}{\eta^4}\left[
(n_1^2+n_2^2+n_3^2+n_4^2)\xi_O+
(2n_1n_2+2n_3n_4)\xi_V+\right.\nonumber\\
&&\left.(2n_1n_3+2n_2n_4)\xi_S+
(2n_1n_4+2 n_2 n_3)\xi_C\right](\frac{i\tau_2}{2})\nonumber\\
{\cal M}&= &\frac{{\cal V}_6}{2}(n_1 +n_2 +n_3 +n_4 )
\int_0^\infty \frac{d \tau_2}{\tau_2^4}\frac{1}{\eta^4}
\hat{\xi}_O(\frac{i\tau_2}{2}+\frac{1}{2}) \quad .
\eea
The expression for $\epsilon= -1$ is similar, it only amounts to
substituting the characters $\xi_0,\xi_V,\xi_S,\xi_C$ with the 
characters
\bea
\tilde{\xi}_O&= &\chi_1+\tilde{\chi}_4\nonumber\\
\tilde{\xi}_V&= &\chi_6+\tilde{\chi}_7\nonumber\\
\tilde{\xi}_S&= &\chi_7+\tilde{\chi}_6\nonumber\\
\tilde{\xi}_C&= &\chi_4+\tilde{\chi}_1 \quad ,
\eea
respectively. Notice that, unlike in the $\epsilon= 1$
case, $\tilde{\xi}_V,\tilde{\xi}_S$ and
$\tilde{\xi}_C$ are massless characters providing with additional
massless matter in the bifundamentals.

For $\epsilon= 1$, the non-supersymmetric choice corresponds to the following
amplitudes
\bea
{\cal K}_{ns}&= &\frac{{\cal V}_6}{2}
\int_0^\infty \frac{d \tau_2}{\tau_2^4}\frac{1}{\eta^4}\left[
\xi_O+\xi_V-\xi_S-\xi_C\right](2i\tau_2)\nonumber\\
{\cal A}_{ns}&= &\frac{{\cal V}_6}{2}
\int_0^\infty \frac{d \tau_2}{\tau_2^4}\frac{1}{\eta^4}\left[
(2 n \bar{n} + 2 m \bar{m})\xi_O+
(n^{2} + \bar{n}^{2} + m^{2} + \bar{m}^{2})\xi_V+\right.\nonumber\\
&&\left.(2n\bar{m}+2\bar{n}m)\xi^{{\prime}}_S+
(2nm+2\bar{n}\bar{m})\xi^{{\prime}}_C\right](\frac{i\tau_2}{2})\nonumber\\
{\cal M}_{ns}&= &\frac{{\cal V}_6}{2}
\int_0^\infty \frac{d \tau_2}{\tau_2^4}\frac{1}{\eta^4}
\left[ (n + \bar{n}) \hat{\xi}_V - (m + \bar{m}) \hat{\xi}^{\prime\prime}_V\right]
(\frac{i\tau_2}{2}+\frac{1}{2}) \quad ,
\eea
where with a prime ($^{\prime}$) and a double prime ($^{\prime\prime}$)
we denote characters corresponding to non-supersymmetric chiral traces 
much in the same way as discussed in Appendix A.
Similar expressions, with $\xi \rightarrow
\tilde\xi$, correspond to the non-supersymmetric choice with $\epsilon= -1$. 
It should be noticed that in this case only $\xi_{V}$ 
($\tilde{\xi_{V}}$) 
enters the
transverse Klein-bottle amplitude.  As already shown, $\xi_{V}$ is
massive, and a model without open-string sector is perfectly 
consistent with both worldsheet and target-space requirements.
\rnc{\Large}{\normalsize}

\end{document}